**Quel jour meurt-on le plus en France ?**

**En 2023, 639 300 personnes sont décédées en France, soit 35 900 personnes de moins qu'en 2022, année de forte mortalité. Sur les vingt dernières années, de 2004 à 2023, le 3 janvier a été le jour le plus meurtrier, tandis que le 15 août a été le jour le moins meurtrier. Les personnes âgées meurent nettement moins souvent en été. Par ailleurs, les décès sont moins fréquents lors des jours fériés et les dimanches. Enfin, le risque de mourir est plus élevé le jour de son anniversaire, surtout pour les jeunes.**

Nathalie Blanpain, Insee focus n°337, octobre 2024

**En 2023, 35 900 décès de moins qu'en 2022**

En 2023, 639 300 personnes sont décédées en France, soit 35 900 personnes de moins qu'en 2022 (-5 %, **figure 1**). Le nombre de décès a reculé après un sommet atteint en 2022, année marquée par cinq vagues de Covid-19, deux épisodes de grippe en avril et en décembre et des périodes de fortes chaleurs en juillet et en août (**figure 2**) [Blanpain, 2023]. En 2023, la baisse du nombre de décès dont la **cause initiale** est la Covid-19 s'est accélérée : -27 000 environ d'après les premiers comptages, après -20 000 en 2022 [Cadillac *et al.*, 2024]. De plus, aucun pic de grippe n'a eu lieu en 2023, puisque celui de l'épidémie de l'hiver 2022-2023 a eu lieu de façon précoce en décembre 2022, alors que celui de l'hiver suivant a eu lieu de façon habituelle en janvier-février 2024 [SpFrance, 2023, 2024]. Enfin, même si des épisodes de forte chaleur et de canicule ont eu lieu en 2023, la mortalité estivale a été nettement plus modérée qu'en 2022.

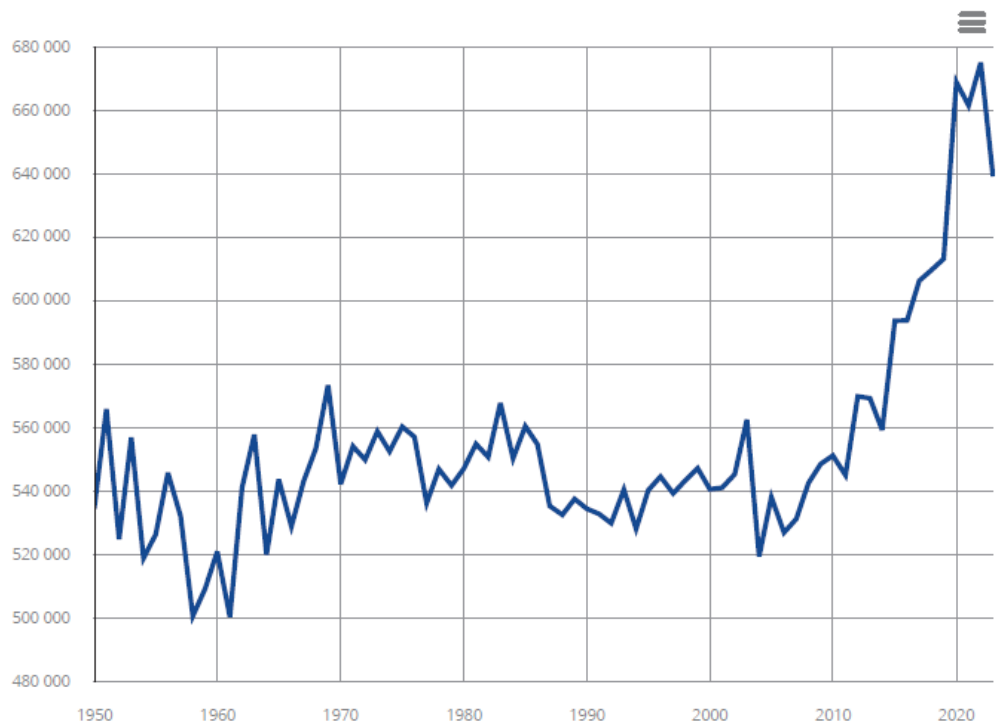

Figure 1 – Nombre de décès par année de 1950 à 2023

Lecture : En 2023, 639 269 personnes sont décédées en France.
Champ : France métropolitaine jusqu'en 1981, France hors Mayotte de 1982 à 2013, France depuis 2014.
Source : Insee, statistiques d'état civil.

Depuis 2011, le nombre de décès a tendance à augmenter du fait de l'arrivée à des âges de forte mortalité des générations nombreuses du baby-boom, nées de 1946 à 1974. La période 2020-2022 est particulière, marquée par une forte mortalité due essentiellement à l'épidémie de Covid-19. En 2023, le nombre de décès est supérieur de 4 % à son niveau pré-pandémique de 2019. Cette hausse s'explique par le vieillissement de la population. Par ailleurs, de 2019 à 2023, l'espérance de vie a progressé deux fois moins vite que sur la dernière décennie : pour les hommes, de 1 mois par an, contre 2 mois par an de 2010 à 2019 ; pour les femmes, de 0,6 mois par an, contre 1 mois par an de 2010 à 2019 [Papon, 2024].

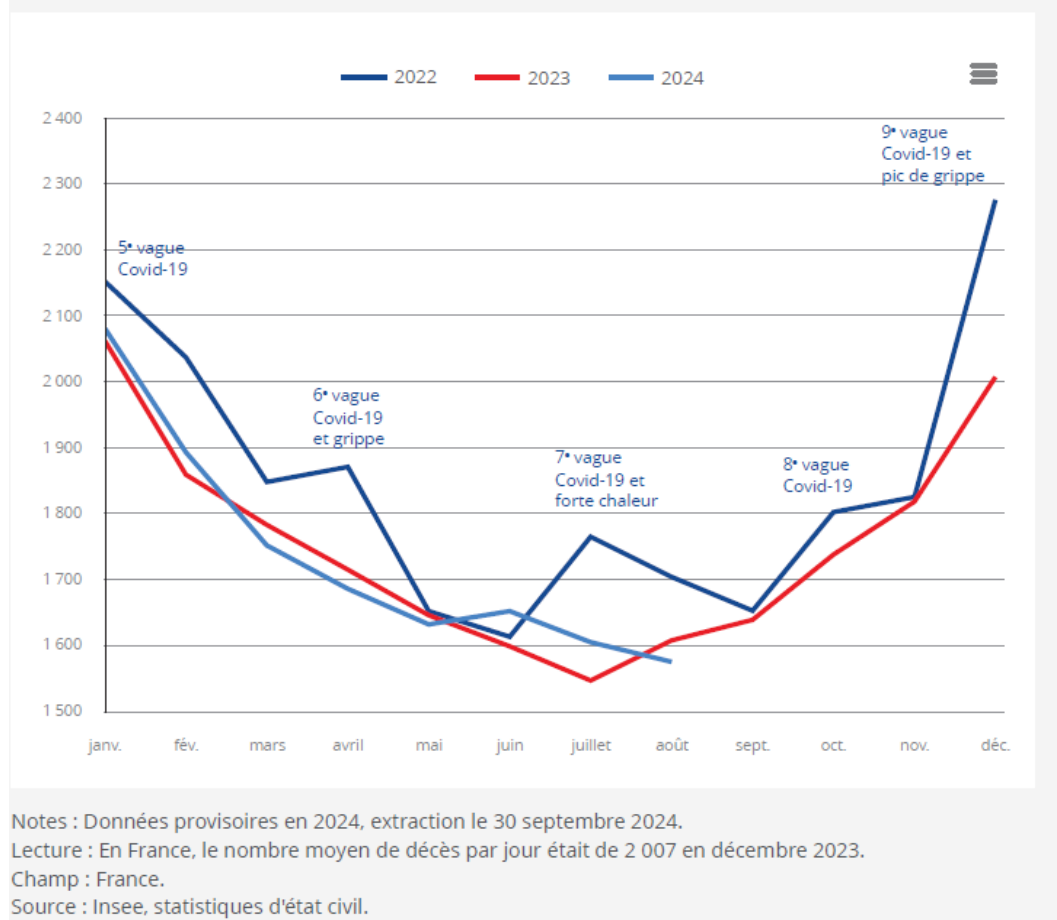

Figure 2 – Nombre moyen de décès par jour selon le mois en 2022, 2023 et 2024

Notes : Données provisoires en 2024, extraction le 30 septembre 2024.
Lecture : En France, le nombre moyen de décès par jour était de 2 007 en décembre 2023.
Champ : France.
Source : Insee, statistiques d'état civil.

**Juin, juillet et août sont les mois les moins meurtriers pour les personnes âgées**

De 2004 à 2023, 1 450 personnes sont décédées en moyenne par jour en août, contre 1 600 sur l'ensemble de la période, soit -9 % (figure 3a). Les mois d'été (juin, juillet, août et septembre) ont été les moins meurtriers, avec une sous-mortalité de -8 % ou -9 % par rapport à l'ensemble de la période, ce qui s'explique notamment par une moindre circulation des virus saisonniers. À l'inverse, les mois d'hiver ont été les plus meurtriers, avec une surmortalité de +9 % en décembre, +14 % en janvier, +12 % en février et +6 % en mars.

Cette saisonnalité est observée chez toutes les personnes âgées de 60 ans ou plus, particulièrement parmi les plus âgées. Ainsi, les personnes de 90 ans ou plus connaissent effectivement une sous-mortalité en juin, juillet et août (-14%) et une surmortalité en janvier (+21 %) et en février (+18 %). L'été, malgré les canicules, est nettement moins meurtrier que l'hiver pour les personnes âgées. En 2003, année de forte canicule, le mois d'août avait été le plus meurtrier de l'année pour les personnes âgées de 90 ans ou plus, avec une surmortalité exceptionnelle de +31 %. Depuis 2004, les personnes de 90 ans ou plus connaissent, au contraire, une sous-mortalité comprise entre -10 % et -21 % en août. Suite à la canicule de 2003, les pouvoirs publics ont en effet mis en place un système de surveillance des vagues de chaleurs et des mesures de prévention à destination notamment des personnes les plus vulnérables.

Les jeunes se distinguent par des décès plus nombreux pendant l'été : les 1-17 ans connaissent une surmortalité en juillet (+11 %) et les 18-29 ans en juin (+3 %), juillet (+7 %) et août (+6 %). Les jeunes décèdent plus fréquemment sur la voie ou dans un lieu public : 12 % des décès des 1-17 ans, 27 % des 18-29 ans, contre 1 % des décès de l'ensemble de la population. Or, les décès dans ces lieux se produisent plus souvent l'été (**figure 3b**) : tous âges confondus, le pic de surmortalité sur la voie ou dans un lieu public survient en juillet (+13 %). Chez les personnes âgées de 30 à 59 ans et les enfants de moins d'un an, les décès varient assez peu selon le mois.

Figure 3a – Écart de mortalité à la moyenne de 2004 à 2023 par mois et par âge

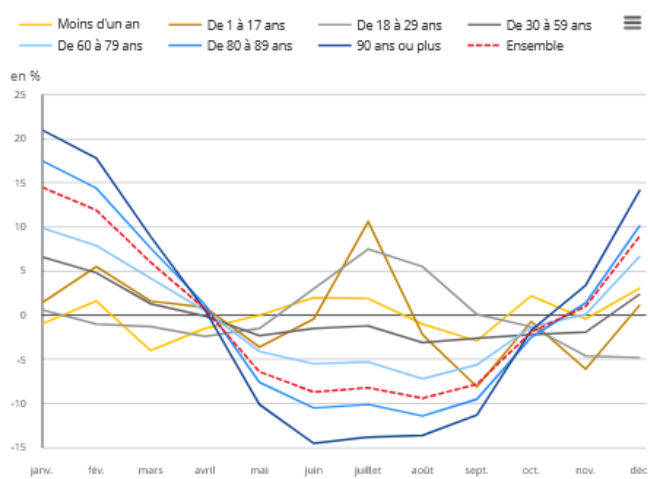

Note : L'âge est révolu.
Lecture : De 2004 à 2023, le nombre moyen des décès des personnes âgées de 1 à 17 ans en juillet était supérieur de 10,6 % au nombre moyen des décès des personnes du même âge sur la période.
Champ : France hors Mayotte jusqu'en 2013, France à partir de 2014.
Source : Insee, statistiques d'état civil.

Figure 3b – Écart de mortalité à la moyenne de 2004 à 2023 par mois et par lieu de décès

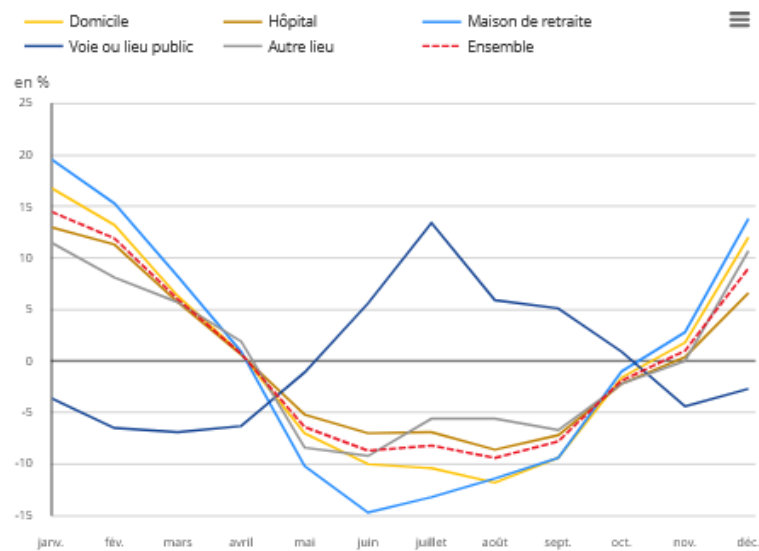

Lecture : De 2004 à 2023, le nombre moyen des décès en maison de retraite en juin était inférieur de 14,7 % au nombre moyen des décès en maison de retraite sur la période.
Champ : France hors Mayotte jusqu'en 2013, France à partir de 2014.
Source : Insee, statistiques d'état civil.

**Le 15 août, jour d'été et férié, est le jour le moins meurtrier de l'année**

De 2004 à 2023, le 3 janvier a été le jour le plus meurtrier de l'année, avec un nombre moyen de décès de 1 900, contre 1 600 sur l'ensemble de la période, soit +19 % (**figure 4**). Le 3 janvier est un jour d'hiver, qui suit les fêtes de fin d'année. Le désir de passer ces fêtes avec des proches, ainsi que celui d'atteindre une nouvelle année pourrait retarder la survenue du décès des personnes en fin de vie et expliquer en partie ce pic. De plus, cette période correspond à une reprise des opérations chirurgicales programmées. À l'opposé, le 15 août a été le jour le moins meurtrier, avec 1 410 décès quotidiens en moyenne, soit -12 % par rapport à l'ensemble de la période. Il s'agit d'un jour d'été, qui est férié.

Or, les décès lors des jours fériés sont moins fréquents. Les décès du 15 août sont inférieurs de 2 % à ceux ayant eu lieu en moyenne lors des 3 jours précédents et suivants. Le jour de Noël, la sous-mortalité est également de -2 % par rapport aux jours adjacents, même si le nombre de décès est supérieur de 10 % au nombre moyen par jour du fait de la saison hivernale. Les décès à l'hôpital diminuent davantage lors des jours fériés : par exemple, -5 % à Noël par rapport aux 3 jours précédents et suivants, contre -2 % tous lieux de décès confondus. Ceci pourrait s'expliquer par des prises en charge moins fréquentes et un moindre nombre d'interventions programmées lors de ces jours de repos habituels.

En revanche, les décès sur la voie ou dans un lieu public augmentent la plupart des jours fériés : +23 % le 1er janvier et +21 % le 14 juillet par rapport aux 3 jours précédents et suivants. Mais ils diminuent à Noël par rapport aux jours adjacents (-18 %) et sont au même niveau le 1er mai (0 %).

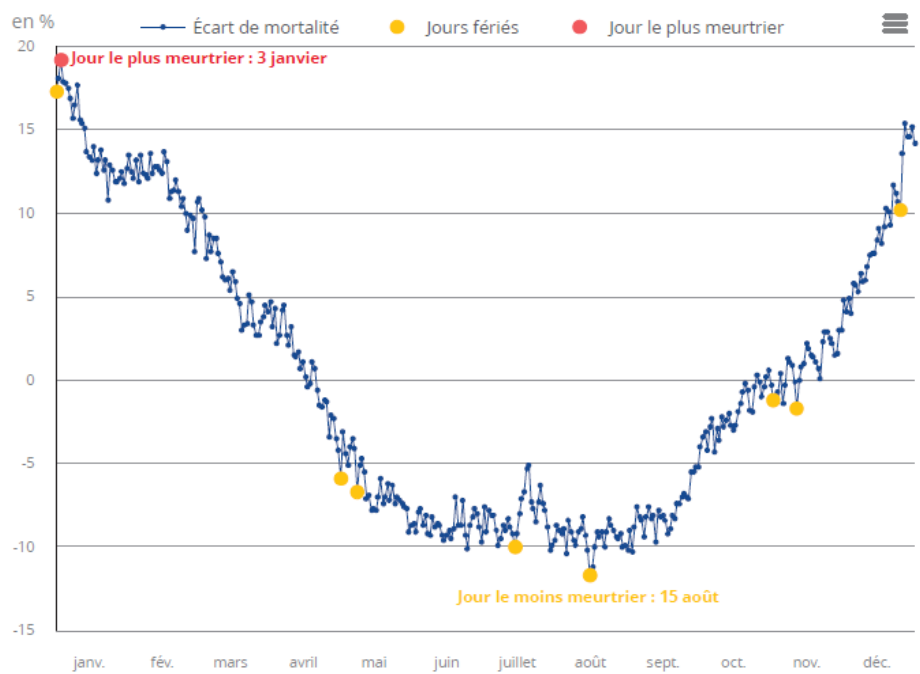

Figure 4 – Écart de mortalité à la moyenne de 2004 à 2023 selon le jour

Note : Seuls les jours fériés à date fixe peuvent être représentés.
Lecture : De 2004 à 2023, le nombre moyen des décès le 15 août était inférieur de 11,7 % au nombre moyen des décès quotidien sur la période.
Champ : France hors Mayotte jusqu'en 2013, France à partir de 2014.
Source : Insee, statistiques d'état civil.

### Le dimanche est le jour de la semaine le moins meurtrier

Tous âges et lieux de décès confondus, le jour le plus meurtrier de la semaine est le mardi, avec 1 620 décès en moyenne, soit une surmortalité de +1 %, tandis que le moins meurtrier est le dimanche, avec 1 550 décès (-3 %), suivi par le samedi, avec 1 580 décès (-1 %) (figure 5). Chez les enfants de moins d'un an, les décès chutent très nettement le dimanche et augmentent le vendredi.

Comme pour les jours fériés, la sous-mortalité le dimanche pourrait s'expliquer par un nombre moins élevé de prises en charge et d'interventions programmées ce jour-là : les décès à l'hôpital diminuent un peu plus que la moyenne le dimanche (-4 %, contre -3 %). Par ailleurs, les visites de la famille ou amis auprès de personnes malades, sans doute plus fréquentes le week-end, pourraient jouer : les décès à domicile ont aussi moins souvent lieu le week-end (-2 % le samedi et le dimanche), mais augmentent le lundi (+4 %). C'est aussi le cas dans les maisons de retraite, mais les différences entre le week-end et le lundi sont moindres.

En revanche, chez les 18-29 ans et de façon moins prononcée chez les 1-17 ans, la mortalité est plus élevée le week-end. Les jeunes sont davantage concernées par les morts accidentelles, sur la route notamment. Chez les 18-29 ans, les décès sur la voie ou dans un lieu public augmentent nettement le samedi et le dimanche, respectivement +32 % et +34 %.

Figure 5 – Écart de mortalité à la moyenne de 2004 à 2023 selon le jour de la semaine et le groupe d'âges

en %

| Jour de la semaine | Jour de la naissance | 1 jour à moins d'un an | De 1 à 17 ans | De 18 à 29 ans | De 30 à 59 ans | De 60 à 69 ans | De 70 à 79 ans | De 80 à 89 ans | 90 ans ou plus | Ensemble |
|---|---|---|---|---|---|---|---|---|---|---|
| Lundi | 6,3 | -8,0 | -2,3 | -4,6 | 2,6 | 1,5 | 0,4 | 0,5 | 0,5 | **0,8** |
| Mardi | 4,5 | 3,0 | -3,5 | -7,9 | 1,1 | 1,7 | 1,7 | 1,0 | 1,1 | **1,2** |
| Mercredi | -0,2 | 4,5 | 0,9 | -7,2 | 0,0 | 0,8 | 0,6 | 0,4 | 0,2 | **0,4** |
| Jeudi | 4,7 | 7,1 | 0,7 | -5,2 | 0,6 | 1,0 | 0,9 | 0,7 | 0,1 | **0,6** |
| Vendredi | 8,5 | 14,5 | -1,4 | 0,1 | 1,0 | 1,8 | 1,3 | 0,5 | 0,0 | **0,8** |
| Samedi | -7,7 | -4,3 | 3,5 | 11,1 | -2,0 | -2,2 | -1,1 | -0,8 | -0,3 | **-1,0** |
| Dimanche | -16,0 | -16,8 | 2,2 | 13,6 | -3,3 | -4,6 | -3,9 | -2,3 | -1,5 | **-2,7** |

Notes : L'âge est révolu. La mortalité est inférieure à -2 % dans les cellules bleues et supérieure à 2 % dans les cellules rouges.
Lecture : De 2004 à 2023, le nombre moyen des décès le dimanche était inférieur de 2,7 % au nombre moyen des décès quotidien sur la période.
Champ : France hors Mayotte jusqu'en 2013, France à partir de 2014.
Source : Insee, statistiques d'état civil.

### Le risque de mourir est plus élevé le jour de son anniversaire, surtout pour les jeunes

Le risque de mourir le jour de son anniversaire est plus élevé que n'importe quel autre jour dans l'année : de 1994 à 2023, la moyenne des décès ce jour-là était supérieure de 6 % à la moyenne de la période (figure 6). Le risque augmente légèrement plus pour les hommes (+7 %, contre +6 % pour les femmes) et fortement pour les jeunes et les adultes d'âge intermédiaire (+15 % de 2 à 17 ans, +21 % de 18 à 29 ans, +21 % de 30 à 39 ans, +13 % de 40 à 49 ans) (figure 7). Les hommes âgés de 18 à 29 ans et de 30 à 39 ans sont ainsi les plus touchés (+24 % de décès le jour de leur anniversaire).

Au-delà de 50 ans, le risque de mourir le jour de son anniversaire reste plus élevé qu'un autre jour, mais de façon moins prononcée : l'écart de mortalité est de +5 % ou +6 % selon les groupes d'âges. Toutefois, les centenaires se distinguent : le nombre moyen des décès le jour des 100 ans est supérieur de 29 % au nombre moyen de décès quotidiens pour les personnes de même âge.

La surmortalité le jour de son anniversaire est particulièrement forte pour les décès sur la voie publique (+14 %) ou au domicile (+12 %). Elle est dans la moyenne en maison de retraite (+7 %) et faible à l'hôpital (+3 %).

Ce phénomène, appelé « syndrome de l'anniversaire » ou « *birthday effect* » en anglais, a été observé dans d'autres pays, comme la Suisse [Ajdacic-Gross *et al.*, 2012] ou les États-Unis [Peña, 2015]. Plusieurs hypothèses sont avancées. En Suisse, les accidents de la route, les chutes et les accidents cardiovasculaires sont plus fréquents ce jour-là, ce qui pourrait s'expliquer par des excès (alcool, fatigue due à la fête…). La dimension psychologique pourrait avoir à la fois un effet négatif et un effet positif. Au Japon, le risque de suicide augmente le jour de son anniversaire [Tetsuya M., Michiko U., 2016]. Cette date symbolique pourrait exacerber un sentiment de tristesse ou de solitude. À l'inverse, le désir d'atteindre le jour de son anniversaire pourrait retarder la survenue du décès des personnes en fin de vie.

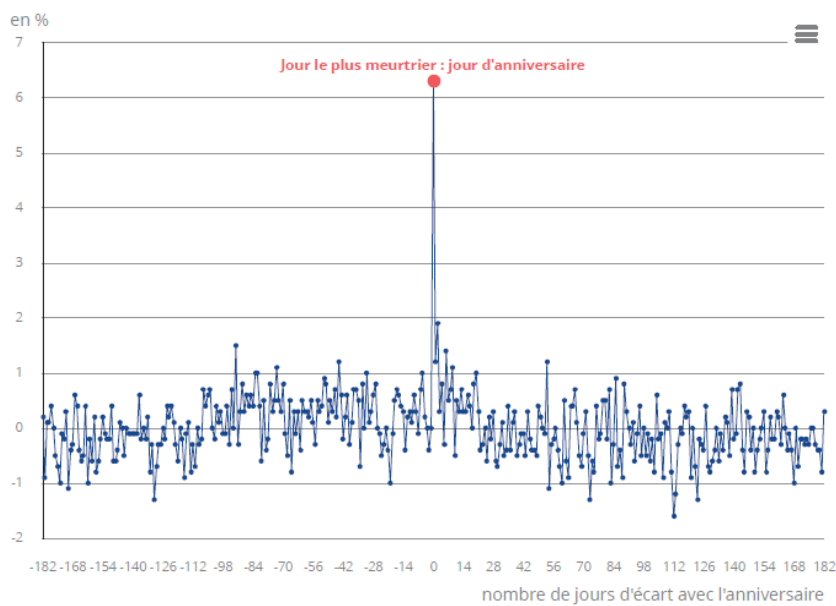

Figure 6 – Écart de mortalité à la moyenne de 1994 à 2023 selon le nombre de jours d'écart avec l'anniversaire

Notes :
- Les personnes âgées de 1 an ou moins sont exclues, car toutes n'ont pas atteint l'âge de leur anniversaire. L'âge est en différence de millésimes.
- Les personnes nées ou décédées un 29 février sont exclues.
Lecture : De 1994 à 2023, le nombre moyen des décès lors du jour d'anniversaire était supérieur de 6,3 % au nombre moyen des décès quotidien sur la période.
Champ : France métropolitaine jusqu'en 1997, France hors Mayotte de 1998 à 2013, France à partir de 2014.
Source : Insee, statistiques d'état civil.

Tableau complémentaire - Écart de mortalité à la moyenne de 1994 à 2023 lors du jour d'anniversaire selon le sexe et l'âge

en %

| Sexe | De 2 à 17 ans | De 18 à 29 ans | De 30 à 39 ans | De 40 à 49 ans | De 50 à 59 ans | De 60 à 69 ans | De 70 à 79 ans | De 80 à 89 ans | 90 ans ou plus | Ensemble |
|---|---|---|---|---|---|---|---|---|---|---|
| Femme | 16 | 14 | 15 | 10 | 3 | 2 | 6 | 7 | 4 | 6 |
| Homme | 15 | 24 | 24 | 15 | 8 | 6 | 5 | 4 | 9 | 7 |
| Ensemble | 15 | 21 | 21 | 13 | 6 | 5 | 6 | 6 | 5 | 6 |

Notes :
- L'âge est en différence de millésime. L'âge révolu lors de l'anniversaire le plus proche de la date de décès n'a pas été retenu, car les personnes décédées après leur anniversaire sont plus âgées que celles décédées avant, ce qui joue sur l'effectif des décès.
- Les personnes âgées de 1 an ou moins sont exclues, car toutes n'ont pas atteint l'âge de leur anniversaire.
- Les personnes nées ou décédées un 29 février sont exclues.
Lecture : De 1994 à 2023, le nombre moyen des décès lors du jour d'anniversaire des personnes âgées de 2 à 17 ans était supérieur de 15 % au nombre moyen des décès quotidien des personnes du même âge sur la période.
Champ : France métropolitaine jusqu'en 1997, France hors Mayotte de 1998 à 2013, France à partir de 2014.
Source : Insee, statistiques d'état civil.

## Sources

Les statistiques d'état civil (dont les décès) sont issues des informations transmises par les mairies à l'Insee. Le code civil oblige en effet à déclarer tout événement d'état civil à un officier d'état civil dans des délais prescrits. L'Insee s'assure de l'exhaustivité et de la qualité des données avant de produire les fichiers statistiques d'état civil.

Le champ de cette étude porte sur l'ensemble des décès ayant eu lieu en France, quel que soit le lieu de résidence. Les données sont définitives jusqu'en 2023 et provisoires en 2024.

## Définitions

La cause initiale de décès est définie par l'Organisation Mondiale de la Santé comme « la maladie ou le traumatisme qui a déclenché l'évolution morbide conduisant directement au décès, ou les circonstances de l'accident ou de la violence qui ont entraîné le traumatisme mortel ».

La surmortalité ou la sous-mortalité lors d'un mois $m$ est le rapport entre le nombre moyen des décès quotidiens d'une population lors de ce mois $m$ sur une période donnée et le nombre moyen des décès quotidiens de cette population sur l'ensemble de cette période.

## Pour en savoir plus